\documentclass[%
 reprint,
 amsmath,amssymb,
 aps,
]{revtex4-2}

\usepackage{graphicx}
 \usepackage[all,cmtip]{xy}
\usepackage{dcolumn}
\usepackage{bm}

\newcommand{\de}{\text{d}}

\newcommand{\genl}[1]{\mathbf{#1}}
\newcommand{\genr}[1]{\overline{\mathbf{#1}}}

\usepackage{color}

\begin{document}


\title{The Tensionless Limit of Pure-Ramond-Ramond AdS3/CFT2}

\author{Alberto Brollo}\email{alberto.brollo@tum.de}
\affiliation{%
Dipartimento di Fisica e Astronomia, Universit\`a degli Studi di Padova, via Marzolo 8, 35131 Padova, Italy.}
\affiliation{%
Zentrum Mathematik, Technische Universit\"at M\"unchen, Boltzmannstra\ss e 3, 85748 Garching, Germany.}

\author{Dennis le Plat}\email{diplat@physik.hu-berlin.de}
\affiliation{Institut f\"ur Mathematik und Physik, Humboldt-Universit\"at zu Berlin, Zum gro{\ss}en Windkanal 2, 12489 Berlin, Germany.}

\author{Alessandro Sfondrini}\email{alessandro.sfondrini@unipd.it}
\affiliation{%
Dipartimento di Fisica e Astronomia, Universit\`a degli Studi di Padova, via Marzolo 8, 35131 Padova, Italy.}
\affiliation{%
Istituto Nazionale di Fisica Nucleare, Sezione di Padova, via Marzolo 8, 35131 Padova, Italy.}
\affiliation{Institute for Advanced Study,
Einstein Drive, Princeton, New Jersey, 08540 USA}

 \author{Ryo Suzuki}%
 \email{rsuzuki.mp@gmail.com}
\affiliation{%
Shing-Tung Yau Center of Southeast University,
No.2 Sipailou, Xuanwu district, Nanjing, Jiangsu, 210096, China.}%

\date{\today}

\begin{abstract}
Despite impressive advances in the AdS3/CFT2 correspondence, the setup involving Ramond-Ramond backgrounds, which is related to the D1-D5 system of branes, remained relatively poorly understood. 
We use the Mirror TBA equations recently constructed by Frolov and Sfondrini to study the spectrum of pure Ramond-Ramond $AdS_3\times S^3\times T^4$ strings. We find that the leading-order contribution to the anomalous dimensions at small tension is due to the gapless worldsheet excitations, \textit{i.e.}\ to the~$T^4$ bosons and their superpartners, whose interactions are nontrivial.

\end{abstract}

\maketitle


\section{Introduction and Summary}
The AdS3/CFT2 correspondence is one of the earliest instances of holography~\cite{Maldacena:1997re}, yet it remains rather mysterious. Even when restricting to simple observables such as the free-string spectrum, little can be computed away from some very special setups.
There exist several maximally supersymmetric $AdS_3$ backgrounds with 16 Killing spinors. Here we consider the simplest, $AdS_3\times S^3\times T^4$.
The background can be supported by a combination of Ramond-Ramond (RR) and Neveu-Schwarz-Neveu-Schwarz (NSNS) fluxes~\cite{Larsen:1999uk,OhlssonSax:2018hgc}, but so far only the setup without RR-fields is well-understood.

Consider strings  on $AdS_3\times S^3\times T^4$ with mixed~flux.
The string tension~$T$ is sourced by the (quantized) NSNS coupling~$k$ and by the (continuous) RR coupling~$g$,
\begin{equation}
\label{eq:tension}
    T=\frac{R^2}{2\pi\alpha'}=\sqrt{g^2+\frac{k^2}{4\pi^2}}\,,\qquad g\geq0,\quad k\in\mathbb{N}_0\,.
\end{equation}where $R$ is the $S^3$~radius. 
This can be read off the bosonic action~$\mathbf{S}_{\text{bos}}$ which is given by a sigma model (SM) term and a Wess-Zumino (WZ) term,
\begin{equation}
    \mathbf{S}_{\text{bos}}=\frac{T}{2}\,\mathbf{S}_{\text{SM}}+\frac{k}{4\pi}\mathbf{S}_{\text{WZ}}\,,\quad
    T\geq0,\quad k\in\mathbb{N}_0\,.
\end{equation}
$T\gg1$ gives the supergravity and semiclassical regimes.
When only NSNS fluxes are present ($g=0$), the worldsheet theory is a level-$k$ supersymmetric Wess-Zumino-Witten (WZW) model~\cite{Giveon:1998ns}, and it can be solved~\cite{Maldacena:2000hw}. Its free spectrum can be easily written in closed-form, and it is hugely degenerate like the spectrum of flat-space strings.
The $g=0$ holographic duals are symmetric-product-orbifold CFTs which are particularly simple at $k=1$~\cite{Giribet:2018ada,Eberhardt:2018ouy} and more subtle for $k\geq2$~\cite{Eberhardt:2021vsx}.

If $g>0$ the worldsheet CFT becomes nonlocal~\cite{Berenstein:1999jq,Cho:2018nfn}, and it is hard to decouple its ghost sector~\cite{Berkovits:1999im,Eberhardt:2018exh}. As a result, the computation of the spectrum and other observables is hard. It is unknown how to describe the holographic duals for generic $g,k$~\footnote{These backgrounds are all related by U-duality, but the map cannot be realised in perturbative string theory.}.
They should be as nontrivial as, for instance, planar $\mathcal{N}=4$ supersymmetric Yang-Mills theory~(SYM) at finite 't~Hooft coupling.

An alternative to the worldsheet-CFT approach is to exploit the classical integrability of the model, which holds for any $g\geq0$ and $k\in\mathbb{N}_0$, as found in~\cite{Cagnazzo:2012se} following~\cite{Henneaux:1984mh,Metsaev:1998it,Bena:2003wd}. By studying the $AdS_3\times S^3\times T^4$ Green-Schwarz (GS) action~\cite{Rahmfeld:1998zn,Pesando:1998wm,Wulff:2013kga} in a suitable lightcone gauge, we may bootstrap the worldsheet S-matrix~\cite{Sfondrini:2014via} --- the same approach used for $AdS_5\times S^5$ and $\mathcal{N}=4$ SYM with remarkable success~\cite{Arutyunov:2009ga,Beisert:2010jr}.
The equations describing the free-string spectrum for pure-RR backgrounds ($k=0$ and any $g>0$) were recently constructed~\cite{Frolov:2021bwp}. This  regime is interesting because it is ``as far as possible'' from the WZW construction,  is directly related to the D1-D5 system of branes in perturbative string theory~\cite{David:2002wn}, and allows for the smallest possible tension~\eqref{eq:tension}.

Our aim is to study this spectrum in the small-tension limit, which is expected to be dual to a weakly-coupled two-dimensional CFT. 
Before presenting our results, we will briefly review the construction of the string lightcone-gauge model~\cite{Borsato:2014exa} and of its ``mirror''~\cite{Arutyunov:2007tc}, describe its particle content, sketch the mirror TBA equations derived in~\cite{Frolov:2021bwp} and discuss their weak-tension limit. We will finally present the spectrum in the tensionless limit and offer some concluding remarks. The detailed derivation of the weak-tension TBA equations and the algorithm for their solution will be presented in an upcoming companion paper~\cite{upcoming}.

\section{Pure-RR Lightcone Gauge-Fixed Model}
The construction of the lightcone-gauge model follows~\cite{Arutyunov:2009ga}. It was performed in~\cite{Borsato:2014hja} by considering the GS action~$\genl{S}$ in the uniform lightcone gauge~\cite{Arutyunov:2005hd}.
The superisometry algebra is $psu(1,1|2)\oplus psu(1,1|2)$. There is also a local $so(4)$ isometry algebra from~$T^4$ which is useful to label the states: Bispinors of $so(4)$ carry indices $A=1,2$, $\dot{A}=1,2$. We denote by $\genl{L}_0$ and $\genr{L}_0$ the $su(1,1)$ Cartan element of either $psu(1,1|2)$ algebra, and by $\genl{J}{}^3$ and $\genr{J}{}^3$ the $su(2)$ ones.  The BPS bound is
\begin{equation}
\label{eq:BPSbound}
    \genl{E}:=\genl{L}_0 -\genl{J}{}^{3}\geq0\,,\qquad
    \genr{E}:=\genr{L}_0 -\genr{J}{}^{3}\geq0\,.
\end{equation}
A point-like string moving along the time direction~$t$ in $AdS_3$ and along a great circle~$\varphi$ in $S^3$~\cite{Berenstein:2002jq}  saturates~\eqref{eq:BPSbound}. It can be used to define the uniform lightcone gauge (supplemented by a lightcone $\kappa$-gauge fixing~\cite{Borsato:2014hja})
\begin{equation}
\label{eq:lcgauge}
    X^+ = \tau,\quad P_-=1,\quad
    X^\pm=\frac{\varphi\pm t}{2},
    \quad P_\mu=\frac{\delta \genl{S}}{\delta \dot{X}^\mu}\,.
\end{equation}
The worldsheet time~$\tau$ is conjugate to the Hamiltonian~$\genl{H}$,
\begin{equation}
    \genl{H}=\genl{E}+\genr{E}\geq0\,,\quad\text{and we define}\quad
    \genl{M}:=\genl{E}-\genr{E}\in\mathbb{Z}\,.
\end{equation}
$\genl{H}$ vanishes on half-BPS states, while $\genl{M}$ is a combination of $AdS_3$ and $S^3$ spins. Due to~\eqref{eq:lcgauge} and to $\kappa$-gauge fixing, only 8 bosons and 8 fermions survive; Furthermore, reparametrization invariance is lost and the model is not Lorenz invariant. 
The surviving symmetries were studied in~\cite{Borsato:2013qpa,Borsato:2014hja}. The algebra undergoes a central extension similarly to Beisert's~\cite{Beisert:2005tm,Arutyunov:2006ak}; The additional central charges must vanish on physical states satisfying the level-matching condition. A perturbative analysis~\cite{Borsato:2014hja} indicates that the eigenvalues of the additional central charges are proportional to the strength of the RR coupling~$g$ introduced in~\eqref{eq:tension}. Algebraic considerations fix the dispersion relation of a single excitation of worldsheet-momentum~$p$~\cite{Borsato:2012ud}
\begin{equation}
\label{eq:stringdispersion}
    H(M,p)=\sqrt{M^2+4h^2\sin^2(\tfrac{p}{2})}\,,
\end{equation}
where $M$ is the eigenvalue of~$\genl{M}$.
Here $h=h(T)$ is an effective coupling depending on the tension~\cite{OhlssonSax:2018hgc}. At strong tension, $h\sim g\sim T$, while $T\to0$ when~$h\to0$. While~$h(T)$ should be determined like in~\cite{Minahan:2009aq,Sundin:2012gc,Gromov:2014eha}, \eqref{eq:stringdispersion} is exact in~$h$ and it reduces to the pp-wave results~\cite{Berenstein:2002jq,Gava:2002xb} in the large-$h$, small-$p$ limit. 
Four bosons on $AdS_3\times S^3$ fit in two irreps with $M=\pm1$, while those on $T^4$ fit in two irreps with $M=0$, labeled by $\dot{A}=1,2$. The eight fermions complete those multiplets. 
Similar algebraic considerations~\cite{Zamolodchikov:1978xm,Arutyunov:2006yd} are sufficient to fix the two-particle S~matrix, which  satisfies the Yang-Baxter equation~\cite{Borsato:2014hja}, up to overall ``dressing'' factors. Closure of the S-matrix bootstrap~\cite{Dorey:1996gd} requires us to introduce appropriate bound states of the fundamental particles, thereby allowing for any~$M\in\mathbb{Z}$~\cite{Borsato:2013hoa}. The dressing factors are constrained by crossing, unitarity and analyticity and were recently proposed in~\cite{Frolov:2021fmj}.

The S~matrix describes the theory on a decompactified worldsheet. To obtain the spectrum of $\genl{H}$ we need the cylinder. Imposing periodic boundary conditions for a $N$-particle state gives the Bethe-Yang equations, schematically
\begin{equation}
\label{eq:BYeqs}
    e^{i p_j L}\prod_{k=1}^N S_{M_j,M_k}(p_jp_k)=-1,\qquad
    j=1,\dots N\,,
\end{equation}
where $L=J{}^3+\bar{J}{}^3$ is the R-charge of the vacuum. (The Bethe-Yang equations are actually more involved and feature ``auxiliary'' excitations because the S-matrix is nondiagonal~\cite{Borsato:2012ss,Seibold:2022mgg}.)
The energy and level-matching conditions read
\begin{equation}
\label{eq:asymptoticEn}
    H=\sum_{j=1}^N H(M_j,p_j)\,,\qquad
    \sum_{j=1}^N p_j=0\,.
\end{equation}
Eqs.~\eqref{eq:BYeqs} and~\eqref{eq:asymptoticEn} are not exact as they neglect finite-size effects~\cite{Ambjorn:2005wa}. These are due to virtual particles wrapping the cylinder and are suppressed at $L\gg1$ by $e^{-LM}$, much like tunneling. As this model features $M=0$ gapless excitations wrapping should be particularly~severe.

\section{Mirror Model and TBA}
Following~\cite{Zamolodchikov:1989cf} we account for wrapping (finite-volume) effects by studying the finite-temperature features of a new model, related to the previous by the exchange of worldsheet time and space,
\begin{equation}
    (\tau,\sigma)\to(-i\tilde{\sigma},-i\tilde{\tau}),\qquad
    (H,p)\to(i\tilde{p},i\tilde{H})\,.
\end{equation}
Because our model is non-relativistic, the dispersion changes drastically~\cite{Ambjorn:2005wa,Arutyunov:2007tc}, from \eqref{eq:stringdispersion} to
\begin{equation}
    \tilde{H}(M,\tilde{p})=2\,\text{arcsinh}\frac{\sqrt{M^2+\tilde{p}^2}}{2h}\,.
\end{equation}
The particle content of the mirror model is similar to that of the original model, and it consists of:
\begin{enumerate}
    \item Gapped excitations with $M=\pm1,\pm2,\dots$,
    \item Gapless excitations with $M=0$ which come in two families, distinguished by~$\dot{A}=1,2$,
    \item Four types of auxiliary particles (labeled by $a=\pm$ and by $A=1,2$) which carry no energy and account for the multiplet structure of the model.
\end{enumerate}
The derivation of the Mirror TBA equations for the ground-state was done in~\cite{Frolov:2021bwp}. The equations are expressed in terms of ``Y-functions'' which give the distribution particles and holes at finite ``temperature''~$1/L$ as a function of $\tilde{p}$ or of a suitable rapidity which we call~$u$. Schematically, they are written in terms of convolutions~\footnote{Repeated indices are summed; $(A,\dot{A})$ indices and the contour choice are suppressed.}
\begin{equation}
\label{eq:tbaeq}
\begin{aligned}
    -\ln Y_{M}(u)=&L\tilde{H}(M,u)-\left[\ln(1+Y_{J})*K_{JM}\right](u)\\
    &-\left[\ln\left(1-\frac{1}{Y_a}\right)*K_{aM}\right](u),
\end{aligned}
\end{equation}
where the kernels are related to the S~matrices by $K_{JM}(u,v)=\tfrac{1}{2\pi i}\tfrac{\de}{\de u}\ln S_{JM}(u,v)$.
For auxiliary particles there is no energy contribution
\begin{equation}
\label{eq:auxTBA}
    \ln Y_a(u)=-\left[\ln(1+Y_{M})*K_{Ma}\right](u).
\end{equation}
Details can be found in~\cite{Frolov:2021bwp}. These ground-state equations can be easily generalized to excited states using a clever analytic continuation~\cite{Dorey:1996re}. During the continuation, some singularities may cross the integration contours. Let $u_j$, $j=1,\dots N$ such that $Y_{M_j}(u_j)=-1$, which we can also write as
\begin{equation}
\label{eq:exactBethe}
    \ln Y_{M_j}(u_j)= i\pi (2\nu_j+1)\,,\qquad \nu_j\in\mathbb{Z}\,.
\end{equation}
Picking up the singularity of $\ln(1+Y_{M_j})$ amounts to adding to the right-hand side of~\eqref{eq:tbaeq} a driving term of the schematic form
\begin{equation}
\label{eq:driving}
    \Delta_{M}({u_j})=\sum_{j=1}^N\ln S_{M_jM}(u_j,u)\,.
\end{equation}
There is also a similar term in~\eqref{eq:auxTBA}.
Finally, the energy for the excited state is given by convolutions over non-auxiliary particles (including both flavors of massless particles)
\begin{equation}
\label{eq:tbaenergy}
    H=-\int \frac{\de u}{2\pi}\frac{\de \tilde{p}_M}{\de u} \ln(1+Y_M)+\sum_{j=1}^NH(M_j,p_j)\,,
\end{equation}
where the last term also comes from the deformation of the contour. The quantization of $p_j=p(u_j)$ follows from imposing~\eqref{eq:exactBethe} on~\eqref{eq:tbaeq}.

\section{Tensionless Limit}
We now write down the excited-state mirror TBA at $h\ll1$. To this end we first worked out the excited-state equations for any $h \ge 0$ (sketched above), and then take the small-$h$ limit.  Interestingly, we find~\cite{upcoming} that the result of this procedure coincides with taking $h\ll1$ in the ground-state equations and applying the contour-deformation trick to those equations.

Let us analyze the Mirror TBA equations~\eqref{eq:tbaeq} as $h\to0$. Let us assume that the convolutions are regular in this limit, which we prove in~\cite{upcoming}. Since
\begin{equation}
    \tilde{H}(M,\tilde{p})= 2\ln\frac{\sqrt{M^2+\tilde{p}^2}}{h}+O(h^2)\,,
\end{equation}
we have that
\begin{equation}
     Y_{M}(u)=h^{2L}\, y_{M}(u)+O(h^{2L+1})\,,\qquad M\neq0\,.
\end{equation}
where $y_{M}(u)$ is regular and $h$-independent. Therefore, the contribution of $M\neq0$ Y-functions is suppressed in the energy~\eqref{eq:tbaenergy} as well is the other TBA equations as~$O(h^{2L})$.
The story is different for $M=0$. Even at small $h$, the small-$|\tilde{p}|$-region of the $M=0$ modes is never suppressed. To better see this, we reparameterize $\tilde{p}_{M=0}$ and $\tilde{H}(0,\tilde{p})$~\cite{Frolov:2021zyc}
\begin{equation}
\label{eq:pHofgamma}
    \tilde{p}=-\frac{2h}{\text{sh}\gamma}\,,\quad
    \tilde{H}=\ln\left(\frac{1+e^\gamma}{1-e^\gamma}\right)^2,\quad\gamma\in\mathbb{R}\,.
\end{equation}
Hence $Y_0(\gamma)$ is finite as $h\to0$ and its integral contributes at $O(h)$ to the energy~\eqref{eq:tbaenergy}, because $\de \tilde{p}/\de\gamma=O(h)$.
The auxiliary functions~$Y_a$ do not enter~\eqref{eq:tbaenergy}, but they are finite as $h\to0$ and  couple to the equations for~$Y_0$; Hence, they cannot be discarded.

\begin{figure*}[t]
    \centering
    \includegraphics[width=\linewidth]{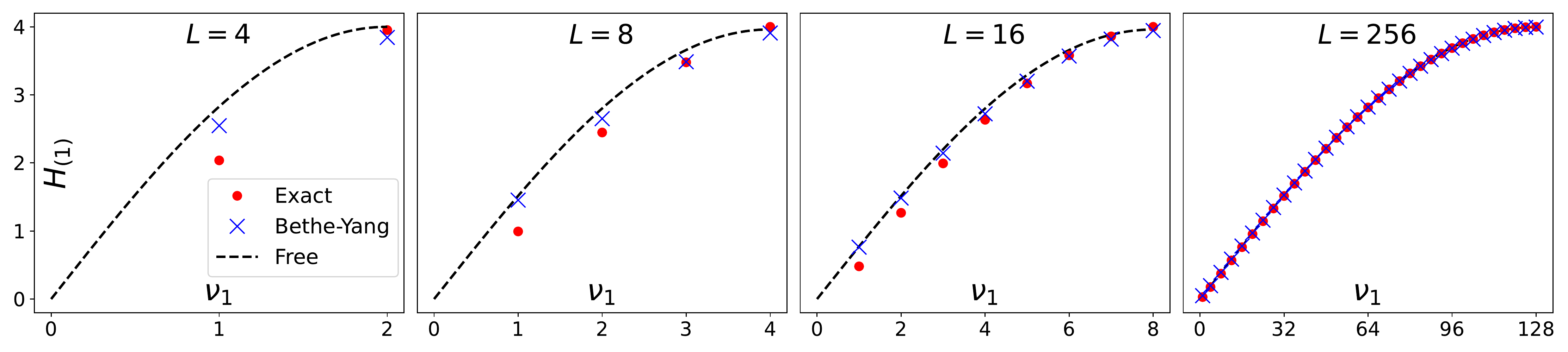}
    \caption{Anomalous dimensions for states with $\nu_1=-\nu_2$. Expanding~$H=H_{(1)}\,h+O(h^2)$, we plot~$H_{(1)}$ for various lengths comparing it with the Bethe-Yang prediction~\eqref{eq:BYeqs} and with the energy of a free model with dispersion~\eqref{eq:asymptoticEn}.}
    \label{fig:plots}
\end{figure*}

Let us compute mirror TBA equations at leading order, \textit{i.e.}~$O(h^0)$ for $Y_0$ with $\dot{A}=1,2$ and for the auxiliary functions $Y_{\pm}$ with $A=1,2$. We consider the case where all excitations~\eqref{eq:exactBethe} are  gapless modes ($M_j=0$) as they are most important at small-$h$. We assume that no extra singularities of~\cite{Arutyunov:2009ax} appear.
We find several remarkable simplifications. Firstly, the number of equations is reduced to just two --- one for the gapless modes, and one for the auxiliary functions. Secondly, the kernels and S-matrices are of difference form (which was the motivation for introducing~$\gamma$ in~\cite{Fontanella:2019baq}). Finally, all kernels reduce to the Cauchy kernel
\begin{equation}
    s(\gamma)=\frac{1}{2\pi i}\frac{\de \ln S(\gamma)}{\de\gamma}\,,\quad
    S(\gamma)=-i\,\text{th}\left(\frac{\gamma}{2}-\frac{i\pi}{4}\right).
\end{equation}
Suppressing the $\gamma$-dependence we write~\footnote{This formula also assumes that $N$ is even, which is the case of interest here.}
\begin{equation}
\label{eq:smallhTBA}
\begin{aligned}
    \ln Y_0&=-L\tilde{H}+\ln[(1+Y_0)^2(1-Y)^4]*s+\Delta_0\,,\\
    \ln Y&=\ln[(1+Y_0)^2]*s+\Delta_0\,,
\end{aligned}
\end{equation}
where the driving term is given by~\eqref{eq:driving} by setting all S~matrices to be $S(\gamma_j-\gamma+\tfrac{i\pi}{2})$. It is easy to see that as a consequence~$Y_0(\gamma_k)=0$. 
We expect~\eqref{eq:exactBethe} to hold in the string region, at $\gamma_k^+:=\gamma_k+\tfrac{i\pi}{2}$:
\begin{equation}
\label{eq:smallhExactBethe}
    i\pi(2\nu_k+1)=-iLp_k-\ln[(1+Y_0)^2(1-Y)^4]*s+\Delta_0,
\end{equation}
where we used~$\tilde{H}(\gamma_k^+)=i p_k$. The energy is finally
\begin{equation}
    H=-\int\frac{\de\gamma}{2\pi} \frac{\de \tilde{p}}{\de\gamma}\ln(1+Y_0)^2+\sum_{j=1}^N H(p_j)\,,
\end{equation}
where we used that $H(p_j)=i \tilde{p}(\gamma_j^+)$.
Note that ${\de \tilde{p}}/{\de\gamma}$ has a pole at $\gamma=0$, \textit{cf.}~\eqref{eq:pHofgamma}. Nonetheless, the integration converges because $Y_0(\gamma)=O(\gamma^{2L})$ around zero due to the $L\tilde{H}$ term in~\eqref{eq:smallhTBA}.
Similarly, the $\tfrac{i\pi}{2}$-shifted Cauchy kernel in~\eqref{eq:smallhExactBethe} is singular in~$\gamma_k$, but the Y-functions vanish there, making all convolutions well-defined.
As it is generally the case for excited-state TBA equations --- with the notable exception of WZW $AdS_3$ backgrounds~\cite{Baggio:2018gct,Dei:2018mfl,Dei:2018jyj} --- it appears impossible to find an analytic solution, and we resort to numerical evaluation.

\begin{figure}[b]
    \centering
    \includegraphics[width=\linewidth]{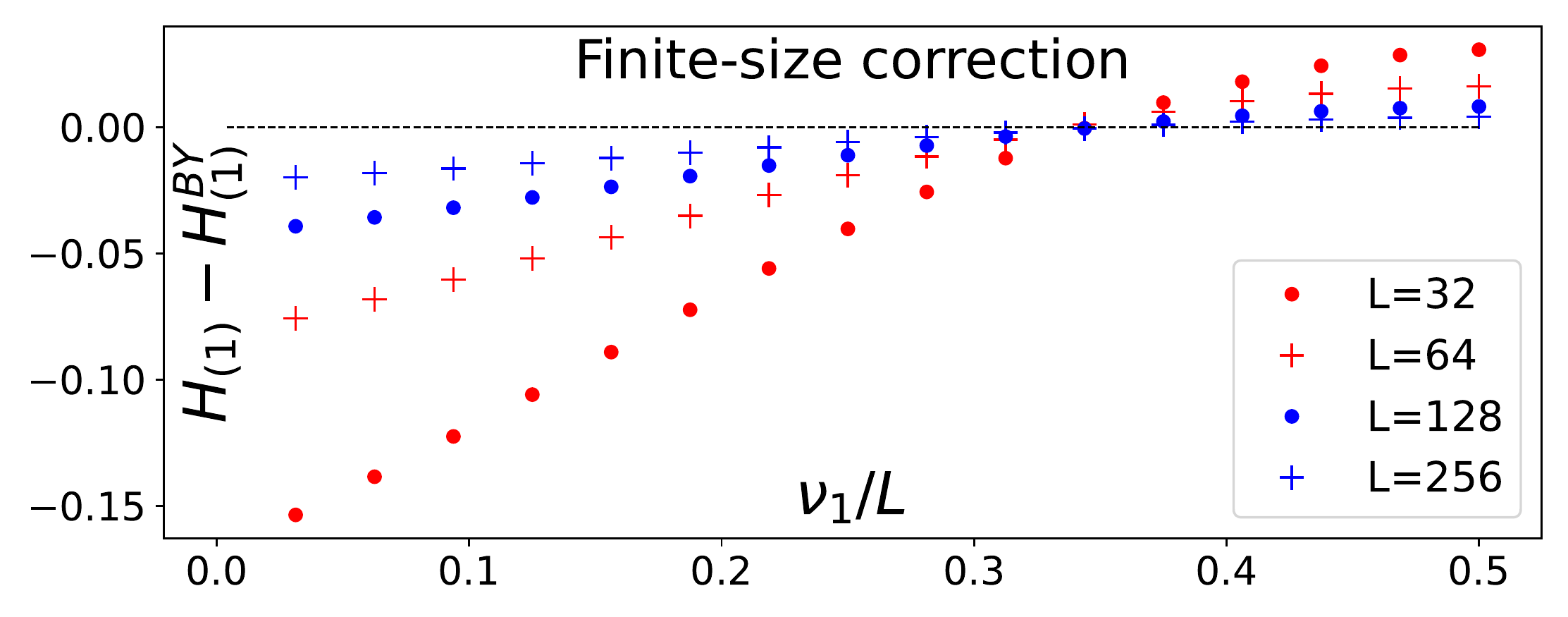}
    \caption{The finite-size correction with respect to the Bethe-Yang prediction decreases roughly as~$1/L$.}
    \label{fig:error}
\end{figure}

\section{Tensionless Spectrum}
Let us summarize the results of the TBA analysis order-by-order in~$h$. At~$O(h^0)$, there is no Y-function contribution to the energy. The only contribution comes from the asymptotic part of the energy~\eqref{eq:asymptoticEn}. Since at this order $H(M,p)=|M|$, see~\eqref{eq:stringdispersion}, the gapped modes contribute with their ``engineering'' dimension, irrespective of their momentum (like in tree-level $\mathcal{N}=4$ SYM) while the gapless ones have zero energy, leading to a glut of degenerate states (like in flat space when $\alpha'=\infty$). At~$O(h^1)$ both the asymptotic energy of gapless modes and their Y-functions contribute---signaling that wrapping occurs as early as possible. This lifts the degeneracy of gapless excitations. The next qualitative difference occurs at~$O(h^{2L})$ when the wrapping of massive states begins contributing to the energy.

\begin{figure}[b]
    \centering
    \includegraphics[width=\linewidth]{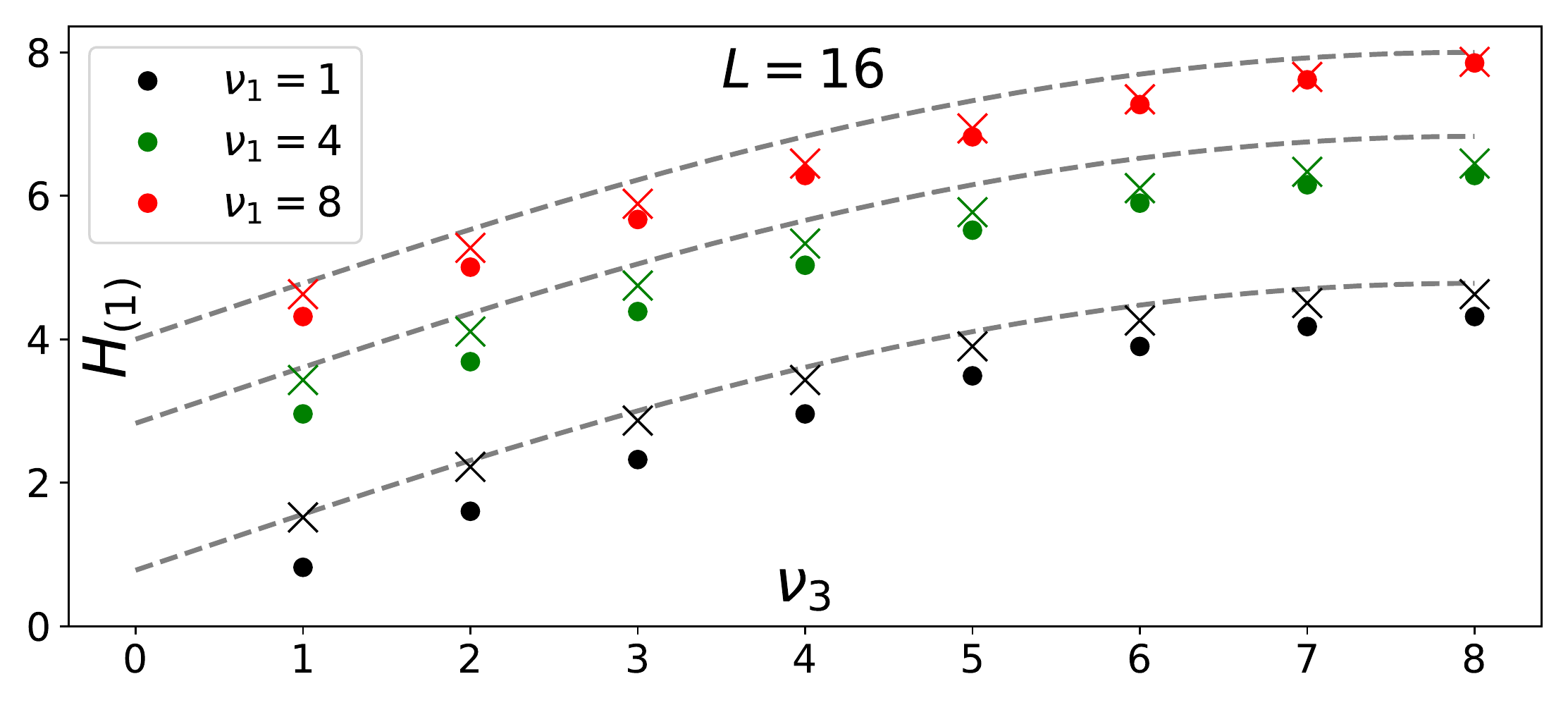}
    \caption{Anomalous dimensions for states with $\nu_1=-\nu_2$ and $\nu_3=-\nu_4$. States with $\nu_1=\nu_3$ are allowed as long as the $su(2)$ labels are $\dot{A}_1\neq \dot{A}_3$; Their energy is regular.}
    \label{fig:N=4}
\end{figure}

We solved~\eqref{eq:smallhTBA} numerically to high precision by iterations from the asymptotic Y-functions (dropping the convolutions). We present the results for the anomalous dimensions at~$O(h^1)$, starting from $N=2$ excitations with $\nu_1=-\nu_2$, which is necessary and sufficient to satisfy level-matching.  Fig.~\ref{fig:plots} shows the anomalous dimensions. We find $L/2$ distinct energies. They are rather well-approximated by the asymptotic result and indeed quite close to the \textit{free} result, \textit{i.e.}~\eqref{eq:asymptoticEn} with~$p_j=2\pi\nu_j/L$. 
Intriguingly, finite-volume corrections do not have a definite sign as a function of~$\nu/L$, and scale like~$1/L$, see also Fig.~\ref{fig:error}. In Fig.~\ref{fig:N=4} we consider $N=4$ states with $\nu_1=-\nu_2$ and $\nu_3=-\nu_4$ (a convenient choice sufficient but not necessary to solve level-matching). 
We see qualitatively similar behavior, and note that the multiparticle energy is not just additive as expected in an interacting model.

\section{Conclusions and Outlook}
We derived the mirror TBA at weak tension for pure-RR $AdS_3\times S^3\times T^4$. It is a simple system of difference-form equations~\eqref{eq:smallhTBA}, whose Y-system can be straightforwardly derived~\footnote{It would be interesting to understand its underlying algebraic structure.}. This TBA describes the spectrum at $O(h^1)$; By contrast, in $AdS_5\times S^5$ (where there are no gapless modes) wrapping effects appear only at~$O(h^{2L})$.

We solved~\eqref{eq:smallhTBA} numerically to high precision. As it happens in the $k=1, g=0$ model, the leading contribution to the energy comes from the $T^4$ modes; However, unlike that case, it is not given by a free theory. It is also different from the $g=0, k\geq2$ spectrum~\cite{Maldacena:2000hw,Baggio:2018gct}, which is of square-root form.
The underlying model does not appear to be a short-range spin-chain either, and it may be described by the gapless sector of the chain investigated in~\cite{OhlssonSax:2014jtq}, which is indeed completely nonlocal. It would be interesting to study that dynamics, which  resembles the recently encountered in four-dimensional $\mathcal{N}=2$ models~\cite{Pomoni:2021pbj}.

Our TBA equations differ from those of~\cite{Bombardelli:2018jkj} as they are of difference form but non-relativistic. Our equations represent the low-tension limit of the spectrum, rather than coming from a low-energy limit of the S-matrix (see also~\cite{Frolov:2023lwd}). Like in~\cite{Bombardelli:2018jkj}, one could extract the central of the dual CFT from the TBA, though this cannot be done with the standard dilogarithm trick precisely because the dispersion relation is non-relativistic. The twisted ground-state energy was recently studied in~\cite{Frolov:2023wji}.

A natural next step is to interpolate from small tension to finite and eventually large tension for a particular set of states, and compare with perturbative results.
A similar computation has been initiated~\cite{Cavaglia:2022xld} using the recently-conjectured ``quantum spectral curve''~\cite{Ekhammar:2021pys,Cavaglia:2021eqr} (QSC). This was done for some states in the gapped sector, for $0<h\lesssim 0.08$. It appears that numerical instabilities make it difficult to extrapolate the QSC beyond that. Furthermore, gapless excitations appear inaccessible in that formalism. It seems however that the TBA equations, while rather cumbersome to treat numerically, do not suffer from similar issues and may be a better numerical testing ground.
Moreover, this would help establish whether the conjectured QSC does indeed match with the mirror TBA as derived from the all-loop S~matrix. This is an important outstanding question that could also be answered through a rigorous derivation of the QSC from the mirror TBA along the lines of~\cite{Bombardelli:2017vhk,Klabbers:2017vtw}.

A more ambitious goal is to extend the mirror TBA to any $g,k$. The integrable structure is modified~\cite{Hoare:2013pma,Hoare:2013lja,Lloyd:2014bsa}, with~\eqref{eq:stringdispersion} becoming
\begin{equation}
    H(M,p)=\sqrt{\left(\tfrac{k}{2\pi}p+M\right)^2+4h^2\sin^2\left(\tfrac{p}{2}\right)}\,,
\end{equation}
with $h=h(T,k)$.
The resulting analytic structure is rather unique, and it so far frustrated the efforts to determine the dressing factor of the theory~\cite{Babichenko:2014yaa}. Expanding on~\cite{Frolov:2021fmj,Frolov:2023lwd} it should be possible to overcome this obstacle.

\begin{acknowledgments}
We thank  Jean-S\'ebastien Caux, Sergey Frolov, Davide Polvara, Stefano Scopa, Fiona Seibold and Dima Sorokin for helpful discussions.

AS~acknowledges support from the European Union -- NextGenerationEU, and from the program STARS@UNIPD, under project ``Exact-Holography'', \textit{A new exact approach to holography: harnessing the power of string
theory, conformal field theory, and integrable models}.
The work of RS is supported by NSFC grant no.~12050410255.
DlP~acknowledges support from the Stiftung der Deutschen Wirtschaft.

DlP, AS, and RS are grateful to the Kavli Institute for Theoretical Physics in Santa Barbara for
hosting them during the Integrable22 workshop, where this work was initiated. AS also thanks the Institute for Advanced Study in Princeton for hospitality during the preparation of this work.

\end{acknowledgments}

\appendix

\bibliographystyle{apsrev4-1} 
\bibliography{apssamp}

\end{document}